\documentclass[preprint,aps,prd,nofootinbib,superscriptaddress]{revtex4}

\usepackage{hyperref}
\usepackage{graphicx}
\usepackage{slashed}
\usepackage{hhline}
\usepackage{subfigure}
\usepackage{epstopdf}

\newcommand{\beq}{\begin{equation}}
\newcommand{\eeq}{\end{equation}}
\newcommand{\beqa}{\begin{eqnarray}}
\newcommand{\eeqa}{\end{eqnarray}}

\newcommand{\nn}{\nonumber}

\newcommand{\Bbar}{\,\overline{\!B}{}}
\newcommand{\Dbar}{\,\overline{\!D}{}}
\newcommand{\Kbar}{\,\overline{\!K}{}}
\def\B0bar{\Bbar{}^0}
\def\D0bar{\Dbar{}^0}
\def\K0bar{\Kbar{}^0}

\usepackage{amsmath}
\usepackage{amsfonts}
\usepackage{amssymb}
\usepackage{latexsym}
\usepackage{graphicx}
\usepackage[english]{babel}
\usepackage{multirow}
\usepackage{float}
\usepackage{url}
\usepackage{hyperref}
\usepackage{slashed}
\usepackage{xcolor}
\usepackage{hyperref}
\usepackage{framed}

\def \beq{\begin{equation}}
\def \eeq{\end{equation}}
\def \bea{\begin{eqnarray}}
\def \eea{\end{eqnarray}}
\def \ba{\begin{array}}
\def \ea{\end{array}}
\def \bmat{\begin{matrix}}
\def \emat{\end{matrix}}
\def \bfra{\begin{framed}}
\def \efra{\end{framed}}

\def \({\left(}
\def \){\right)}
\def \[{\left[}
\def \]{\right]}
\def \<{\left\langle}
\def \>{\right\rangle}
\def \lp{\left|}
\def \rp{\right|}

\def \l.{\left.}
\def \r.{\right.}

\def \bma{\(\bmat}
\def \ema{\emat\)}

\def \nn{\nonumber}
\def \nl{\nn \\}

\def \tth{\frac{2}{3}}
\def \hf{\frac{1}{2}}

\def \s{\sqrt{2}}

\def \sx{\sqrt{6}}

\def \cA{{\cal A}}
\def \ocA{{\overline\cA}}
\def \cB{{\cal B}}

\def \cH{{\cal H}}

\def \oq{\overline{q}}

\def \ou{\overline{u}}
\def \od{\overline{d}}
\def \os{\overline{s}}
\def \oc{\overline{c}}
\def \ob{\overline{b}}

\def \oK{\overline{K}^0}

\def \al{\alpha}

\def \ga{\gamma}

\def \De{\Delta}

\def \la{\lambda}

\def \eff{{\rm eff}}

\def \CP{{\rm CP}}

\def \Im{{\rm Im}}

\def \cH{{\cal H}}

\def\bwt{\begin{widetext}}
\def\ewt{\end{widetext}}

\def \ok{\overline{K}^0}

\newcommand \braket[3]{\< #1 \rp #2 \lp #3 \>}

\newcommand \gr[1]{{\bf #1}}
\newcommand \grb[1]{{\bf\overline{#1}}}

\allowdisplaybreaks

\begin{document}
\preprint{WSU-HEP-1711}

\title{\boldmath Hadronic decays of $B_c$ mesons with flavor $SU(3)_F$ symmetry}

\author{Bhubanjyoti Bhattacharya\footnote{bbhattach@ltu.edu}}
\affiliation{Department of Natural Sciences,\\
Lawrence Technological University, Southfield, MI, 48075, USA}
\affiliation{Department of Physics and Astronomy,
Wayne State University, Detroit, MI 48201, USA}
\author{Alexey A. Petrov\footnote{apetrov@wayne.edu}}
\affiliation{Department of Physics and Astronomy,
Wayne State University, Detroit, MI 48201, USA}

\begin{abstract}
We study implications of a recent observation of non-leptonic $B^+_c\to D^0 K^+$ decay
and a bound on $B^+_c\to D^0 \pi^+$ transition on CP-violating asymmetries in $B_c$
decays. In the U-spin symmetry limit, we derive a relation between the CP-asymmetries
in the $B^+_c\to D^0 K^+$ and $B^+_c\to D^0 \pi^+$ channels and the corresponding
branching ratios. We also derive several relations between non-leptonic $B_c$ decays
into the final states with $D$ mesons in the flavor $SU(3)_F$ limit.
We point out that a combined study of  $SU(3)_F$ amplitudes in these 
decays can be used to constrain the angle $\gamma$ of the 
Cabibbo-Kobayashi-Maskawa (CKM) matrix.
\end{abstract}


\maketitle

\section{Introduction}\label{Intro}

The $B_c$ meson contains a $\ob$ and a $c$ quark, making it a open-flavored
meson with two heavy quarks. Just like its heavy-light cousins $B^0_{d,s}$ and $B^\pm$,
it decays via weak interactions in the Standard Model (SM). However, unlike
those states, decays of the $B_c^+$ meson involve weak decays of either heavy quark
($\ob$ or $c$). Moreover, since tree-level decays of the charm quark involve
transitions between first and second generation quarks, the CKM matrix
elements that come into play ($V_{cs}$ or $V_{cd}$) are large. In contrast,
tree-level decays of the bottom quark involve transitions from the third
generation to the second (or the first) generation and the associated CKM
matrix elements ($V_{ub}$ or $V_{cb}$) are suppressed by one (or more) powers
of the Wolfenstein parameter ($\lambda \sim 0.2$) \cite{Wolfenstein:1983yz}.
Therefore, tree-level weak decays of the $B^+_c$ meson are dominated by the 
$c\to s$ transition. However, unlike in decays of the charm quark where penguin 
amplitudes are suppressed by the small Wilson coefficients of the corresponding 
operators, in bottom quark decays, the Wilson coefficients of the penguin operators 
are quite large. These facts make $B_c$ mesons interesting laboratories for 
simultaneous studies of $b$ and $c$ quark decays.

Recently, the LHCb collaboration observed clear evidence for a $B_c$ decay
that proceeds through the decay of the $b$ quark. Using data with integrated
luminosity of 3.0 $fb^{-1}$ and center-of-mass energies of 7 and 8 TeV, the
LHCb collaboration observed the decay $B^+_c\to D^0 K^+$ at 5.1 $\sigma$
significance \cite{Aaij:2017kea}. The same search also found no evidence for
the U-spin related decay $B^+_c\to D^0\pi^+$. This is a rather interesting result, as
the color-allowed tree-level amplitude in $B^+_c\to D^0\pi^+$ is larger than that for
$B^+_c\to D^0 K^+$ by a factor of $|V_{ud}/V_{us}|\sim 5$.
These observations prompted the collaboration to conclude that the decays are
dominated by weak-annihilation and penguin diagrams rather than the color-favored
tree diagrams.

The LHCb observation of the hadronic $B^+_c$ decay has opened a gateway to
further studies of $B^+_c$ decays, as future measurements in other decay channels
can be expected \cite{Patrignani:2016xqp}. In this letter we study various implications
of the LHCb observation of $B_c$ mesons via the decays of a $b$ quark, particularly
for the observation of CP-violating asymmetries in $B_c$ decays. Using a U-spin symmetry
relation between the decay amplitudes of $B^+_c\to D^0\pi^+$ and $B^+_c\to D^0K^+$
we derive a model-independent relation between CP-violating asymmetries in these channels.
We then generalize our considerations to discuss relationships between different
$B^+_c$ nonleptonic decays under flavor $SU(3)_F$ symmetry.
Compared to similar studies in heavy-light mesons $B_q$, flavor $SU(3)_F$
relations are simpler for $B_c$ mesons, owing to the fact that $B_c$ state
is an $SU(3)_F$ singlet. Our studies complement earlier predictions for branching
ratios and CP asymmetries for various $B^+_c$ decays using perturbative QCD
\cite{Rui:2011qc} and other techniques \cite{Liu:1997hr,Ivanov:2006ni,Choi:2009ym}.

This letter is organized as follows. In Section \ref{U-spin} we show that U-spin
symmetry leads to a convenient relationship between branching ratios and CP
asymmetries in several $B_c^+$ meson decays. Section \ref{SU3} includes a discussion
of more general relationships between decays from a flavor-$SU(3)_F$ symmetry perspective.
We conclude in Section \ref{conclusions}.

\section{U-spin symmetry relations in $B^+_c$ decays}
\label{U-spin}

Decays of the $B^+_c$ meson with $\Delta b = 1$ and $\Delta s = 0(1)$ proceed
through the quark-level transitions $\ob\to\oq\od(\os) q$, where $q = u, c$ at tree-level.
Additionally, $B^+_c$ decays are mediated by transition operators that represent
gluonic-penguin operators ($\ob\to\od(\os)\sum_q \oq q$, with $q = u, d, s, c$) and
electro-weak penguin operators ($\ob\to\od(\os)\sum_q e_q\oq q$, with $q = u, d, s, c$.)
All these operators generate decay amplitudes that are symmetric under the
interchange of $d$ and $s$ quarks (or U-spin symmetry) due to the fact that $u, c, b,$
and $(\od d + \os s)$ are all singlets under U-spin. Thus, there are $\Delta s = 0(1)$
$B^+_c$ decay pairs that are related by U-spin symmetry, implying amplitude-level
relationships between pairs of decays obtained through the exchange $s \leftrightarrow
d$, such as $B_c^+\to D^0\pi^+(K^+)$ and $B_c^+\to D^+ K^0(D_s^+\ok)$. It is interesting
that those U-spin relations between the observed branching ratios also imply relations
between CP-violating asymmetries on those channels.

Let us define CP-asymmetries in the conventional way,
\beq\label{AcpDef}
A_\CP(B_c \to f) = \frac{\Gamma(B_c^+ \to f) - \Gamma(B_c^- \to \bar f)}{\Gamma(B_c^+ \to f) +
\Gamma(B_c^- \to \bar f)},
\eeq
where $\Gamma(B_c \to f)$ is a partial width for $B_c$ transition to the final state $f$. Note that
experimentally reported branching ratios ${\cal B}(B_c \to f)$ are usually averaged over the
CP-conjugated states,
\beq\label{Branch}
{\cal B}(B_c \to f) = \frac{1}{2 \Gamma}\left[\Gamma(B_c^+ \to f) + \Gamma(B_c^- \to \bar f)\right],
\eeq
where $1/\Gamma = \tau = (0.507 \pm 0.009)$ ps is a total lifetime of a $B_c$ state \cite{Patrignani:2016xqp}.

The decays $B^+_c\to D^0\pi^+$ and $B^+_c\to D^0K^+$ are related by the $U$-spin symmetry.
Indeed, we can write the transitions amplitudes for those decays
\bea
\cA(B^+_c\to D^0\pi^+) &=& V^*_{cb}V^{}_{cd}A^c_{d} + V^*_{ub}V^{}_{ud}A^u_d
~,~~
\nonumber \\
\ocA(B^-_c\to \overline{D}^0\pi^-) &=& V^{}_{cb}V^*_{cd}A^c_d + V^{}_
{ub}V^*_{ud}A^u_d ~,~~
\nonumber \\
\cA(B^+_c\to D^0K^+) &=& V^*_{cb}V^{}_{cs}A^c_s + V^*_{ub}V^{}_{us}A^u_s
~,~~ \\
\ocA(B^-_c\to \overline{D}^0K^-) &=& V^{}_{cb}V^*_{cs}A^c_s + V^{}_
{ub}V^*_{us} A^u_s ~,~~
\nonumber
\eea
where CP-conjugate amplitudes were obtained by changing the sign of the
CP-violating phases in the Cabibbo-Kobayashi-Maskawa matrix elements.
Note that we already used unitarity of the CKM matrix to eliminate top-quark-related
combination $V_{tb}^* V_{tq}$ for $q=s,d$. This implies that hadronic matrix
elements contain the corresponding penguin contributions. One
can then construct the following differences in squared amplitudes,
\bea
\lp\cA(B^+_c\to D^0\pi^+)\rp^2 - \lp\ocA(B^-_c\to \overline{D}^0\pi^-)
\rp^2 &=& 4~\Im[V^*_{cb}V^{}_{cd}V^{}_{ub}V^{*}_{ud}]~\Im[A^{c*}
_dA^u_d] ~,~~
\nonumber \\
\lp\cA(B^+_c\to D^0K^+)\rp^2 - \lp\ocA(B^-_c\to \overline{D}^0K^-)
\rp^2 &=& 4~\Im[V^*_{cb}V^{}_{cs}V^{}_{ub}V^{*}_{us}]~\Im[A^{c*}
_sA^u_s] ~.~~
\eea
U-spin implies relationships between the amplitudes: $A^{c,u}_d = A^{c,u}_s$.
Further, unitarity of the CKM matrix leads to the following relationship
\bea
\Im\[V^{*}_{cb}V^{}_{cd}V^{}_{ub}V^{*}_{ud}\] &=& - \Im\[V^{*}_{cb}V^{}_{cs}
V^{}_{ub}V^{*}_{us}\] ~.~~
\eea
Using the above relationships we can show that
\bea
\lp\cA(B^+_c\to D^0\pi^+)\rp^2 - \lp\ocA(B^-_c\to \overline{D}^0\pi^-)
\rp^2 &=& - \lp\cA(B^+_c\to D^0K^+)\rp^2 + \lp\ocA(B^-_c\to \overline
{D}^0K^-)\rp^2. \ \
\eea
Now converting the amplitudes to partial widths and using Eqs.~(\ref{AcpDef})
and (\ref{Branch}) we obtain
\bea
\dfrac{A_\CP(B^+_c\to D^0\pi^+)}{A_\CP(B^+_c\to D^0K^+)} &=& -~\dfrac{p^*
_{\pi^+}}{p^*_{K^+}}~\dfrac{\cB(B^+_c\to D^0K^+)}{\cB(B^+_c\to D^0\pi^+)}~,~~
\eea
where $p^*_M$ represents the magnitude of the three-momentum of the daughter
particles in the rest frame of the decaying $B^+_c$ meson for the decay process
$B^+_c \to D^0 M$. Note that while we have established this relation for the U-spin
related pair of decays $B^+_c\to D^0 \pi^+(K^+)$, the same relation also
exists for other U-spin related pairs such as $B_c^+\to D^+ K^0(D_s^+\ok)$.

The LHCb collaboration recently observed the process $B_c^+\to D^0\pi^+$ with
a significance of 5.1 standard deviations \cite{Aaij:2017kea}. However, since
the absolute production rate for $B_c^+$ at LHCb is unknown, the measured
observable includes a normalizing factor $f_c/f_u$ that compares the production
rate of $B_c^+$ to that of $B^+_u$ using the decay constants of the two mesons,
\bea
R_{D^0K} &=& \dfrac{f_c}{f_u} \times \cB(B^+_c\to D^0K^+) ~=~ \(9.3^{+2.8}_{-2.5}
\pm 0.6\) \times 10^{-7} ~.~~
\eea
At the same time LHCb did not observe any events in the $D^0\pi^+$ channel
thereby putting an upper bounds on the corresponding branching ratio,
\bea
R_{D^0\pi} &=& \dfrac{f_c}{f_u} \times \cB(B^+_c\to D^0\pi^+) ~<~ 3.9 \times
10^{-7} {\rm~at~95~\%~c.l.}~.~~
\eea
The above limit is a direct consequence of no $B^+_c \to D^0\pi^+$ signal events
being seen at the LHCb. Combining the above limits and using $p^*_{\pi^+}/p^*_{K
^+} = 1.008$ we find the following bound on the ratio of CP-violating asymmetries,
\bea\label{ExpBoundAcp}
\left|\dfrac{A_\CP(B^+_c\to D^0\pi^+)}{A_\CP(B^+_c\to D^0K^+)}\right|  &\gtrsim& 2.4 ~,~~
\eea
which is independent of the unknown ratios of the production rates $f_c/f_u$.
Note that the signs of CP-violating asymmetries for $B^+_c \to D^0\pi^+$ and
$B_c^+\to D^+ K^0(D_s^+\ok)$ are {\it opposite} of each other. Note also that
earlier predictions of CP-violating asymmetries \cite{Rui:2011qc,Liu:1997hr}
explicitly violate our model-independent bound Eq.~(\ref{ExpBoundAcp}).

\section{Flavor $SU(3)_F$ analysis of $B_c^+$ decays}\label{SU3}

The U-spin relations derived in the previous section can be generalized by using
full flavor $SU(3)_F$. Here we shall concentrate on the relations among different two-body
decays of the $B_c$ meson to the final state containing an open-flavor heavy and a
light pseudoscalar meson $M=\pi, \eta, K$, such as $B_c \to DM$ and $B_c \to BM$.

We use standard $SU(3)_F$ representations of the initial and final state mesons.
The initial $B_c$ meson transforms as a singlet under $SU(3)_F$, while $B_q$ and $D$ mesons
containing light quarks ($u, d, s$) form triplets and can be represented as row vectors,
\beq
B_i = \left(B^-, B^0_d, B^0_s\right), \quad
D_i = \left(D^0, D^+, D_s\right).
\eeq
The octet of pseudoscalar mesons $M$ formed by the light quarks ($u, d, s$)
can be represented by a $\gr{3} \times \gr{3}$ matrix $M^i_j$ where the upper index represents the rows
(quarks) and the lower index represents the columns (antiquarks). In this notation, the
matrix $M^i_j$ can be expressed as,
\bea
M &=& \bma \dfrac{\pi^0}{\s} + \dfrac{\eta_8}{\sx} & \pi^+ & K^+ \\
          \pi^- & -\dfrac{\pi^0}{\s} + \dfrac{\eta_8}{\sx} & K^0 \\
            K^- & \ok & -\sqrt{\dfrac{2}{3}}\eta_8 \ema ~.~~
\eea
For simplicity we ignore $\eta-\eta^\prime$ mixing and assume $\eta=\eta_8$ from now on.
In order to write $SU(3)_F$ relations for the $B_c$ decay matrix elements we need to specify
representations of the effective Hamiltonians.

\subsection{$B_c$ decays via $b$-quark decay}

The low energy effective Hamiltonian governing $\De b = 1, \De c = 0$ weak decays of the
$\ob$ quark are well known \cite{BBL, GPY},
\bea
\cH &=& \dfrac{4 G_F}{\s} \sum\limits_{q = d,s}\left[V^*_{ub}V^{}_{uq} [C_1 (\ob_L\ga^\mu
u_L)(\ou_L\ga_\mu q_L) + C_2(\ob_L\ga^\mu q_L)(\ou_L\ga_\mu u_L)] - V^*_{tb}V^{}_{tq}
\sum_{i=3}^{10} C_i Q^{(q)}_i\right] \label{eq:bquarkH}~.~~
\eea
The terms corresponding to the coefficients $C_{1,2}$ are generally referred to
as the ``tree'' part, while the rest of the Hamiltonian involving the coefficients
$C_i ~(i ~=~ 3 ~-~ 10)$ are collectively known as the ``penguin'' part. There are
four ``gluonic'' penguin operators $(i ~=~ 3 ~-~ 6)$ and four ``electro-weak''
penguin operators $(i ~=~ 7 ~-~ 10)$. The precise form of these penguin operators
is \cite{BBL},
\bea\label{EffOp}
O^{(q)}_3 ~=~ \sum\limits_{q'' = u,d,s}(\ob_L\ga^\mu q_L)(\oq''_L\ga_\mu q''_L)
~,~~ &~~&
O^{(q)}_4 ~=~ \sum\limits_{q'' = u,d,s}(\ob_L\ga^\mu q''_L)(\oq''_L\ga_\mu q_L) ~,~~
\nonumber \\
O^{(q)}_5 ~=~ \sum\limits_{q'' = u,d,s}(\ob_L\ga^\mu q_L)(\oq''_R\ga_\mu q''_R)
~,~~ &~~&
O^{(q)}_6 ~=~ \sum\limits_{q'' = u,d,s}(\ob_L\ga^\mu q''_L)(\oq''_R\ga_\mu q_R)~,~~ \\
O^{(q)}_7 ~=~ \dfrac{3}{2}\sum\limits_{q'' = u,d,s}(\ob_L\ga^\mu q_L)e_{q''}(\oq''
_R\ga_\mu q''_R) ~,~~ &~~&
O^{(q)}_8 ~=~ \dfrac{3}{2}\sum\limits_{q'' = u,d,s}(\ob_L\ga^\mu q''_L)e_{q''}
(\oq''_R\ga_\mu q_R) ~,~~
\nonumber \\
O^{(q)}_9 ~=~ \dfrac{3}{2}\sum\limits_{q'' = u,d,s}(\ob_L\ga^\mu q_L)e_{q''}(\oq''
_L\ga_\mu q''_L) ~,~~ &~~&
O^{(q)}_{10} ~=~ \dfrac{3}{2}\sum\limits_{q'' = u,d,s}(\ob_L\ga^\mu q''_L)e_{q''}
(\oq''_L\ga_\mu q_L) ~.~~
\eea
In the above operators $q_{L(R)}$ represents left-handed (right-handed) quark fields
and $e_{q}$ represents the electric charge of the quark $q$.

Written in the form of Eq.~(\ref{EffOp}), the operators $O_i$ mix under flavor $SU(3)_F$
transformations. A standard approach\footnote{Another approach to decomposing
decay amplitudes in terms of $SU(3)_F$ matrix elements can be found in
\cite{Bhattacharya:2014eca}} (pioneered in \cite{Shifman:1975tn} for kaon decays)
is to decompose the operators according to different representations of $SU(3)_F$ 
\cite{Savage:1989ub}.

The light quarks ($u, d, s$) transform as a triplet under flavor $SU(3)_F$. The tree part of the
Hamiltonian proportional to $V^*_{ub}V^{}_{uq}$ is made up of four-quark operators of the
form $(\ob q_1)(\oq_2 q_3)$ where $q_i$ represents a light quark. These operators transform
as a $\gr{3}\times\grb{3}\times\gr{3}\equiv\gr{15} + \grb{6} + \gr{3} + \gr{3}$ of $SU(3)_F$.

The part of the Hamiltonian proportional to $V^*_{cb}V^{}_{cq}$ gets contributions from both
trees and penguins. Ignoring contributions from the electroweak penguin operators
$O^{(q)}_{7-10}$, this part transforms as a triplet of $SU(3)_F$.

In what follows, we use the notations introduced by Savage and Wise \cite{Savage:1989ub}
to carry out group-theoretic calculations. Let us first consider $\De s = 0$ transitions.
The triplet Hamiltonian with quantum numbers of $(\ob c)(\oc d)$ can be represented as $H_i = (0, 1, 0)$.
The Hamiltonian with the quantum numbers of $(\ob u)(\ou d)$ is obtained by considering its $SU(3)_F$
decomposition. Here $H(\grb{3})^i \equiv (0, 1, 0)$ is a three-component vector, while $H(\grb{15})^{ij}_k$ and
$H(\gr{6})^{ij}_k$ are traceless three-index tensors that are symmetric ($\grb{15}$) and antisymmetric
($\gr{6}$) on their upper indices \cite{Savage:1989ub}. The non-zero elements of these tensors
are \cite{Savage:1989ub}
\bea
H(\grb{15})^{12}_1 ~=~ H(\grb{15})^{21}_1 ~=~ 3 ~,~~
H(\grb{15})^{22}_2 ~=~ -2 ~,~~
H(\grb{15})^{32}_3 ~=~ H(\grb{15})^{23}_3 ~=~ -1 ~,~~
\nonumber \\
H(\gr{6})^{12}_1 ~=~ -H(\gr{6})^{21}_1 ~=~ 1 ~,~~
H(\gr{6})^{32}_3 ~=~ -H(\gr{6})^{23}_3 ~=~ -1 ~.~~~~~~~~~~~~
\eea
The effective Hamiltonian then takes the following form,
\bea
H_\eff &=& V^*_{cb}V^{}_{cd} \ H^{\oc c\od} + V^*_{ub}V^{}_{ud} \ H^{\ou
u\od} ~,~~ \mbox{where}
\\
H^{\oc c\od} &=& \al B_c H^i M^j_i D_j ~,~~
\nonumber \\
H^{\ou u\od} &=& A_{(\grb{3})} B_c H(\grb{3})^i M^j_i D_j + A_{(\gr{
6})} B_c H(\gr{6})^{ij}_k M^k_i D_j + A_{(\grb{15})} B_c H(\grb{15})
^{ij}_k M^k_i D_j ~.
\nonumber
\eea
The coefficients $\al$, and $A_{(\gr r)}$ respectively represent the reduced
matrix elements from the corresponding group-theoretic operators.
The amplitude for every $B^+_c \to M D$ transition can then be expressed as
$\braket{DM}{H_\eff}{B_c^+}$. We list the results for the corresponding rates in
the upper part of Table \ref{tab:dsratelistI}.
\begin{table}[!htb]
\caption{List of $B_c^+ \to MD$ decay rates for $\De s = 0, 1$ processes
\label{tab:dsratelistI}}
\begin{center}
\begin{tabular}{|c|c|} \hline\hline
Process & Rate \\ \hline\hline
~~~$B^+_c \to D^+\pi^0$~~~ & ~~~$\hf\lp V^*_{cb}V^{}_{cd}~\al + V^*_{ub}V^{}_
{ud}(A_{\grb{3}} - A_{\gr{6}} - 5 A_{\grb{15}})\rp^2$~~~ \\
$B^+_c \to D^0\pi^+$ & $\lp V^*_{cb}V^{}_{cd}~\al + V^*_{ub}V^{}_{ud}
(A_{\grb{3}} - A_{\gr{6}} + 3 A_{\gr{15}})\rp^2$ \\
$B^+_c \to D^+_s\ok$ & $\lp V^*_{cb}V^{}_{cd}~\al + V^*_{ub}V^{}_{ud}
(A_{\grb{3}} + A_{\gr{6}} - A_{\gr{15}})\rp^2$ \\
$B^+_c \to D^+\eta$ & $\frac{1}{6}\lp V^*_{cb}V^{}_{cd}~\al + V^*_{ub}
V^{}_{ud}(A_{\grb{3}} + 3 A_{\gr{6}} + 3 A_{\grb{15}})\rp^2$ \\ \hline
$B^+_c \to D^+K^0$ & $\lp V^*_{cb}V^{}_{cs}~\al + V^*_{ub}V^{}_{us}(A_
{\grb{3}} + A_{\gr{6}} - A_{\grb{15}})\rp^2$ \\
$B^+_c \to D^0K^+$ & $\lp V^*_{cb}V^{}_{cs}~\al + V^*_{ub}V^{}_{us}(A_
{\grb{3}} - A_{\gr{6}} + 3 A_{\grb{15}})\rp^2$ \\
$B^+_c \to D^+_s\pi^0$ & $2\lp V^*_{ub}V^{}_{us}(A_{\gr{6}} + 2 A_{\grb
{15}})\rp^2$ \\
$B^+_c \to D^+_s\eta$ & $\tth\lp V^*_{cb}V^{}_{cs}~\al + V^*_{ub}V^{}_{
us}(A_{\grb{3}} - 3 A_{\grb{15}})\rp^2$ \\ \hline\hline
\end{tabular}
\end{center}
\end{table}

The consideration of $\De s = 1$ transitions arising from weak-interaction
Hamiltonians with the quantum numbers of $(\ob c)(\oc s)$ and $(\ob u)(\ou
s)$ is quite similar. The triplet Hamiltonian with quantum numbers of $(\ob
c)(\oc s)$ is $H_i = (0, 0, 1)$ \cite{Savage:1989ub}. The Hamiltonian with
the quantum numbers of $(\ob u)(\ou s)$ is once again decomposed into irreducible
representations of $SU(3)_F$. Here, $H(\grb{3})^i \equiv (0, 0, 1)$ is a
three-component vector, while $H(\grb{15})^{ij}_k$ and $H(\gr{6})^{ij}_k$ are
traceless three-index tensors that are symmetric ($\grb{15}$) and antisymmetric
($\gr{6}$) on their upper indices. The non-zero elements of these tensors are
\bea
H(\grb{15})^{13}_1 ~=~ H(\grb{15})^{31}_1 ~=~ 3 ~,~~
H(\grb{15})^{33}_3 ~=~ -2 ~,~~
H(\grb{15})^{23}_2 ~=~ H(\grb{15})^{32}_2 ~=~ -1 ~,~~
\nonumber \\
H(\gr{6})^{13}_1 ~=~ -H(\gr{6})^{31}_1 ~=~ 1 ~,~~
H(\gr{6})^{23}_2 ~=~ -H(\gr{6})^{32}_2 ~=~ -1 ~.~~~~~~~~~~~~
\eea
As before, the effective Hamiltonian takes the form,
\bea
H_\eff &=& V^*_{cb}V^{}_{cs} \ H^{\oc c\os} + V^*_{ub}V^{}_{us} \ H^{\ou u\os}
~,~~ \mbox{with}
\\
H^{\oc c\os} &=& \al B_c H^i M^j_i D_j ~,~~
\nonumber \\
H^{\ou u\os} &=& A_{(\grb{3})} B_c H(\grb{3})^i M^j_i D_j + A_{(\gr{6})}
B_c H(\gr{6})^{ij}_k M^k_i D_j + A_{(\grb{15})} B_c H(\grb{15})^{ij}_k M^k
_i D_j ~.~~
\nonumber
\eea
We list the results for the rates of $\De s = 1, ~B^+_c \to DM$ decays in
the bottom part of Table \ref{tab:dsratelistI}.
From the results given in Table \ref{tab:dsratelistI} we can once again see
the relations between the ratios of CP-violating asymmetries and
the corresponding branching ratios,
\bea
\dfrac{A_\CP(B^+_c\to D^0\pi^+)}{A_\CP(B^+_c\to D^0K^+)} &=& -~\dfrac{p^*
_{\pi^+}}{p^*_{K^+}}~\dfrac{\cB(B^+_c\to D^0K^+)}{\cB(B^+_c\to D^0\pi^+)}~,~~
\nonumber \\
\dfrac{A_\CP(B^+_c\to D_s\ok)}{A_\CP(B^+_c\to D^+K^0)} &=& -~\dfrac{p^*
_{D_s}}{p^*_{D^+}}~\dfrac{\cB(B^+_c\to D^+K^0)}{\cB(B^+_c\to D_s\ok)}~.
\eea

Table \ref{tab:dsratelistI} lists eight decay processes, the branching ratio and
direct CP asymmetry for each of which can in principle be measured. Thus there
are 16 observable quantities that can provide information about these decays.
However, flavor $SU(3)_F$ symmetry introduces 2 relationships between these
observables, so that not all of them are independent. If we fix one overall
CP-even phase, then there are a total of 7 hadronic parameters that can fully
characterize the 4 complex $SU(3)_F$ matrix elements represented by $\al$, and
$A_{(\gr r)}$. Clearly, enough information will be available from the branching
ratios and CP asymmetries of the 8 decay processes to determine the 7 hadronic
parameters using a phenomenological fit. However, most of these decay rates
and CP asymmetries have not yet been measured. Future measurements of these
observables will make it possible to study the hadronic parameters in further
detail.

In addition to the seven hadronic parameters mentioned above, there is one CP-odd
phase (the CKM angle $\ga$, the phase of the CKM matrix element $V_{ub}$),
which can be included as an unknown parameter in a fit. Such a fit can provide
a way of obtaining information about the parameter $\ga$, independent of those
commonly used to study it. This further emphasizes the need to measure the
observables mentioned above.

\subsection{$B_c$ decays via $c$-quark decay}

Similar $SU(3)_F$ relations can be obtained for the two-body $B^+_c$
decays to a $B$ meson and a light pseudoscalar meson. Unlike the decays considered
in the previous sub-section, these decays are $\De b = 0, \De c = 1$ transitions.
The Hamiltonian governing quark-level transitions in which the $c$ quark decays,
is similar to that governing the decay of the $b$ quark. However, in this the
penguin operators are dynamically much more suppressed, because the heaviest quark
that can run in the penguin loop is now a $b$ quark. Thus, the largest contributions
come simply from the tree-level operators corresponding to $O_1$, and $O_2$.
\begin{table}[!htb]
\caption{List of $B_c^+ \to MB$ decay rates for CF, SCS, and DCS processes
\label{tab:dsratelistII}}
\begin{center}
\begin{tabular}{|c|c|} \hline\hline
Process & Rate \\ \hline\hline
~~~$B^+_c \to B^0_s\pi^+$~~~ & $\lp V^*_{cs}V^{}_{ud}(2 A^c_{\gr{15}} - A^c_{\grb{6}})\rp^2$ \\
$B^+_c \to B^+\oK$     & $\lp V^*_{cs}V^{}_{ud}(2 A^c_{\gr{15}} + A^c_{\grb{6}})\rp^2$ \\ \hline \hline
$B^+_c \to B^0_s K^+$  & $\lp V^*_{cs}V^{}_{us}(A^c_{\gr{15}} + A^c_{\gr{3}})\rp^2$ \\
$B^+_c \to B^+ \eta$   & ~~~$\frac{1}{6}\lp V^*_{cs}V^{}_{us}(3A^c_{\gr{15}} - A^c_{\gr{3}})\rp^2$ ~~~\\
$B^+_c \to B^0_d\pi^+$ & $\lp V^*_{cd}V^{}_{ud}(A^c_{\gr{15}} + A^c_{\gr{3}})\rp^2$ \\
$B^+_c \to B^+\pi^0$   & $\hf\lp V^*_{cd}V^{}_{ud}(3A^c_{\gr{15}} - A^c_{\gr{3}})\rp^2$ \\ \hline \hline
$B^+_c \to B^0_d K^+$  & $\lp V^*_{cd}V^{}_{us}(2 A_{\gr{15}} - A_{\grb{6}})\rp^2$ \\
$B^+_c \to B^+ K^0$    & $\lp V^*_{cd}V^{}_{us}(2 A_{\gr{15}} + A_{\grb{6}})\rp^2$ \\ \hline\hline
\end{tabular}
\end{center}
\end{table}

Tree-level decay amplitudes where the $c$ quark decays can be classified into three
categories based on the CKM matrix elements that are involved. Cabibbo favored (CF)
amplitudes are proportional to the combination $V_{cs}^*V_{ud}^{}$. Singly Cabibbo
suppressed (SCS) amplitudes are proportional to $V_{cd}^*V_{ud}$ or $V_{cs}^*V_{us}$.
Doubly Cabibbo suppressed (DCS) amplitudes are proportional to the combination $V^*_
{cd}V_{us}$. Compared to the CF amplitudes, the SCS (DCS) amplitudes are suppressed
by one (two) power(s) of the Wolfenstein parameter $\la$. The hadronic
part of the decay amplitudes can be expressed in terms of group-theoretic amplitudes
$A^c_{(\gr{r})}$ where the superscript $c$ denotes that these amplitudes represent decays
of the $c$ quark. The amplitudes for these processes are given in Table
\ref{tab:dsratelistII}.

We find that the amplitudes listed in Table \ref{tab:dsratelistII} satisfy the following
relations,
\bea
\lp\frac{A(B^+_c\to B^0_s\pi^+) + A(B^+_c\to B^+\oK)}{V^*_{cs}V^{}_{ud}}\rp ~=~ 
\lp\frac{A(B^+_c\to B^0_dK^+) + A(B^+_c\to B^+K^0)}{V^*_{cd}V^{}_{us}}\rp \nl
~=~ \lp\frac{A(B^+_c\to B^0_sK^+) + \sx A(B^+_c\to B^+\eta)}{V^*_{cs}V^{}_{us}}\rp \nl
~=~ \lp\frac{A(B^+_c\to B^0_d\pi^+) + \s A(B^+_c\to B^+\pi^0)}{V^*_{cd}V^{}_{ud}}\rp~.~~
\eea

The LHCb collaboration has observed the decay process $B^+_c\to B^0_s\pi^+$ \cite{Aaij:2013cda}.
However, the other decays have not yet been measured.

\section{Conclusions}\label{conclusions}

We considered implications of a recent observation of non-leptonic $B^+_c\to D^0 K^+$ decay and
a bound on $B^+_c\to D^0 \pi^+$ transition on CP-violating asymmetries in $B_c$
decays.We derived two model-independent relations between the CP-asymmetries
in the $B^+_c\to D^0 K^+$ and $B^+_c\to D^0 \pi^+$ and $B^+_c\to D_s\ok$ and
$B^+_c\to D^+K^0$ channels and the corresponding branching rations. We also derived
several relations between non-leptonic $B_c$ decays into the final states with $D$ and $B$ mesons
in the flavor $SU(3)_F$ limit.

While we concentrated on the final states with the pseudoscalar mesons, the same relations
also hold for the pseudoscalar-vector final states. In particular, the results listed in
Tables \ref{tab:dsratelistI} and \ref{tab:dsratelistII} hold exactly for the $D^*M$ and $B^*M$
final states with trivial substitutions $D(B) \to D^*(B^*)$ and $\alpha \to \beta$ and
$A_{(\gr r)}^{(c)} \to B_{(\gr r)}^{(c)}$, where $\beta$ and $B_{(\gr r)}^{(c)}$ are reduced matrix
elements for the effective Hamiltonians describing $B_c \to M D^*(B^*)$ transitions. Only upper
bounds for such decay rates are currently available \cite{Patrignani:2016xqp}.

\acknowledgments
We would like to thank Gil Paz and David London for useful discussions. This work has been supported
in part by the U.S. Department of Energy under contract DE-SC0007983. AAP would like to thank the
University of Siegen for hospitality where part of this work was performed and supported in part by
Comenius Guest Professorship.

\clearpage


\begin{thebibliography}{99}

\bibitem{Wolfenstein:1983yz}
  L.~Wolfenstein,
  Phys.\ Rev.\ Lett.\  {\bf 51}, 1945 (1983).

\bibitem{Aaij:2017kea}
  R.~Aaij {\it et al.} [LHCb Collaboration],
  Phys.\ Rev.\ Lett.\  {\bf 118}, no. 11, 111803 (2017)

\bibitem{Patrignani:2016xqp}
  C.~Patrignani {\it et al.} [Particle Data Group],
  Chin.\ Phys.\ C {\bf 40}, no. 10, 100001 (2016).

\bibitem{Rui:2011qc}
  Z.~Rui, Z.~T.~Zou and C.~D.~Lu,
  Phys.\ Rev.\ D {\bf 86}, 074008 (2012)

\bibitem{Liu:1997hr}
  J.~F.~Liu and K.~T.~Chao,
  Phys.\ Rev.\ D {\bf 56}, 4133 (1997).

\bibitem{Ivanov:2006ni}
  M.~A.~Ivanov, J.~G.~Korner and P.~Santorelli,
  Phys.\ Rev.\ D {\bf 73}, 054024 (2006)

\bibitem{Choi:2009ym}
  H.~M.~Choi and C.~R.~Ji,
  Phys.\ Rev.\ D {\bf 80}, 114003 (2009)
  
\bibitem{BBL}
  G.~Buchalla, A.~J.~Buras and M.~E.~Lautenbacher,
  Rev.\ Mod.\ Phys.\  {\bf 68}, 1125 (1996)

\bibitem{GPY}
  M.~Gronau, D.~Pirjol and T.~M.~Yan,
  Phys.\ Rev.\ D {\bf 60}, 034021 (1999)
  Erratum: [Phys.\ Rev.\ D {\bf 69}, 119901 (2004)]
  
\bibitem{Shifman:1975tn}
  M.~A.~Shifman, A.~I.~Vainshtein and V.~I.~Zakharov,
  Nucl.\ Phys.\ B {\bf 120}, 316 (1977).

\bibitem{Savage:1989ub}
  M.~J.~Savage and M.~B.~Wise,
  Phys.\ Rev.\ D {\bf 39}, 3346 (1989)
  Erratum: [Phys.\ Rev.\ D {\bf 40}, 3127 (1989)].

\bibitem{Bhattacharya:2014eca}
  B.~Bhattacharya, M.~Gronau, M.~Imbeault, D.~London and J.~L.~Rosner,
  Phys.\ Rev.\ D {\bf 89}, no. 7, 074043 (2014)

\bibitem{Aaij:2013cda}
  R.~Aaij {\it et al.} [LHCb Collaboration],
  Phys.\ Rev.\ Lett.\  {\bf 111}, no. 18, 181801 (2013)

\end{thebibliography}
\end{document}